\documentclass[aps, prd, twocolumn,tightenlines, notitlepage, superscriptaddress, nofootinbib, preprintnumbers, floatfix,showkeys,11pt,altaffilletter]{revtex4-2}

\usepackage{upgreek}

\usepackage[official]{eurosym}
\usepackage[normalem]{ulem}
\usepackage{amstext}
\usepackage{graphicx}
\graphicspath{{figures/}}
\usepackage{url}
\usepackage{color}
\usepackage{ulem}
\usepackage[version=4]{mhchem}
\usepackage[utf8]{inputenc}
\usepackage{fontawesome}
\usepackage{yfonts}

\pdfoutput=1
\usepackage{textcomp}
\usepackage{comment}
\usepackage{yfonts}
\usepackage{epsfig,amsfonts,mathrsfs,graphicx,color,slashed,multirow}
\usepackage{latexsym,graphicx,slashed,color,enumerate,url,cancel,gensymb}
\usepackage{textcomp}

\usepackage[x11names]{xcolor}
\usepackage[colorlinks,pdfstartview=FitV,breaklinks=true]{hyperref}
\usepackage{booktabs}
\usepackage{adjustbox}
\usepackage{textgreek} 
\usepackage{caption, subcaption} 
\usepackage{float}

\usepackage{lmodern}
\usepackage{ae,aecompl}
\usepackage{appendix}


\definecolor{vdrgreen}{rgb}{0.0, 0.6, 0.0}

\definecolor{persiangreen}{rgb}{0.0, 0.65, 0.58}
\definecolor{mediumpersianblue}{rgb}{0.0, 0.4, 0.65}

\makeatletter
    \newcommand{\colorboxed}[3][white]{\fcolorbox{#2}{#1}{\m@th$\displaystyle#3$}}
\makeatother
\DeclareMathOperator{\diag}{diag}

\AtBeginDocument{\hypersetup{citecolor=mediumpersianblue,linkcolor=mediumpersianblue,urlcolor=mediumpersianblue}}
\usepackage{appendix}

\begin{document}

\title{{\LARGE SN1987A bounds on neutrino quantum decoherence}}

\author{Christoph A. Ternes}
\email{christoph.ternes@lngs.infn.it}
\affiliation{Istituto Nazionale di Fisica Nucleare (INFN), Laboratori Nazionali del Gran Sasso, 67100 Assergi, L’Aquila (AQ), Italy
}

\author{Giulia Pagliaroli}
\email{giulia.pagliaroli@lngs.infn.it}
\affiliation{Istituto Nazionale di Fisica Nucleare (INFN), Laboratori Nazionali del Gran Sasso, 67100 Assergi, L’Aquila (AQ), Italy
}

\author{Francesco L. Villante}
\email{francesco.villante@lngs.infn.it}
\affiliation{Istituto Nazionale di Fisica Nucleare (INFN), Laboratori Nazionali del Gran Sasso, 67100 Assergi, L’Aquila (AQ), Italy
}
\affiliation{Department of Physical and Chemical Sciences, University of L’Aquila, 67100 L’Aquila, Italy
}

\begin{abstract}
We obtain stringent bounds on neutrino quantum decoherence from the analysis of SN1987A data. We show that for the decoherence model considered here, which allows for neutrino-loss along the trajectory, the bounds are many orders of magnitude stronger than the ones that can be obtained from the analysis of data from reactor neutrino oscillation experiments or neutrino telescopes.
\end{abstract}
\maketitle

\section{Introduction}
\label{sec:intro}

The interpretation of current neutrino oscillation data relies on the so-called three-neutrino paradigm, where neutrinos propagate as superpositions of different mass or flavor eigenstates. During propagation, each component mass eigenstate evolves with a different frequency, thus resulting in the phenomenon of flavor conversion.
This quantum superposition is maintained even over macroscopic distances because neutrinos interact only very weakly with ordinary matter and since they do not decay.  In general, the study of neutrino oscillations considers the neutrino to be isolated from its environment, and the oscillation effects to be coherent. 
However, neutrinos may loose their quantum superposition, for example if the neutrino system interacts with a stochastic environment, thus resulting in a loss of coherence and in the damping of neutrino oscillation probabilities -- a phenomenon known as neutrino decoherence.
Indeed, the possible loss of flavor-coherence can also occur within standard quantum mechanics due to neutrino wave-packet separation~\cite{Giunti:1991sx,Kiers:1995zj,Ohlsson:2000mj,Beuthe:2002ej,Giunti:2003ax,Blennow:2005yk,Farzan:2008eg,Kayser:2010pr,Naumov:2013uia,Jones:2014sfa,Akhmedov:2019iyt,Grimus:2019hlq,Naumov:2020yyv,deGouvea:2020hfl,deGouvea:2021uvg,Akhmedov:2022bjs,Krueger:2023skk,deGouvea:2024syg}. However, here we consider nonstandard decoherence effects which can have different physical origins.

We work in the commonly used framework of open quantum systems~\cite{Gago:2000qc,Benatti:2000ph,Lisi:2000zt,Benatti:2001fa,Breuer:2002pc,Morgan:2004vv,Anchordoqui:2005gj,Fogli:2007tx,Farzan:2008zv, Oliveira:2013nua, Oliveira:2016asf, Jones:2014sfa,  BalieiroGomes:2016ykp, Coelho:2017zes, Coelho:2017byq, Carpio:2017nui,   Coloma:2018idr, Carpio:2018gum, Carrasco:2018sca, Buoninfante:2020iyr, Gomes:2020muc, Ohlsson:2020gxx,Stuttard:2020qfv,Stuttard:2021uyw,Banerjee:2022slh,Barenboim:2024zfi} assuming that neutrino propagation can be affected by the interaction of the neutrino subsystem with the environment. 
The fluctuating nature of space-time (the quantum foam) in a quantum theory of gravity is a commonly cited potential source of a stochastic background that might produce neutrino decoherence effects~\cite{Hawking:1976ra,Ellis:1983jz,Giddings:1988cx,Addazi:2021xuf,Stuttard:2020qfv} via interactions of neutrinos with virtual black holes~\cite{Stuttard:2020qfv}. These Planck-length scale black holes form from extreme fluctuations in the space-time foam and almost immediately evaporate over Planck-time scales. Neutrinos might encounter these virtual black holes along their trajectories which could induce a loss of quantum information. 

Ref.~\cite{Stuttard:2020qfv} discusses several models of neutrino quantum decoherence, in particular the so-called state-selection, phase-perturbation, and neutrino-loss models, which will be briefly introduced below. 
Experimental searches and future sensitivities for neutrino quantum decoherence have been performed with a variety of facilities using different neutrino sources~\cite{Lisi:2000zt,Fogli:2007tx,deOliveira:2013dia,BalieiroGomes:2016ykp,Coelho:2017zes,BalieiroGomes:2018gtd,Carpio:2018gum,Coloma:2018idr,deHolanda:2019tuf,Gomes:2020muc,JUNO:2021ydg,Barenboim:2024wdn,KM3NeT:2024jji}. These searches nearly always focus on state-selection and phase-perturnation models. The most stringent limits on decoherence parameters with positive energy dependence ($\Gamma\propto E^{n}, n>0$, see discussion below) have been obtained using atmospheric neutrinos observed at the IceCube Neutrino Observatory~\cite{ICECUBE:2023gdv}, while for negative dependence they come from reactor neutrino experiments~\cite{DeRomeri:2023dht}. 
Using these experiments, the bounds for state-selection and phase-perturbation are nearly identical and it can be expected that the bound for neutrino-loss is of similar strength, too. 
It should be noted, however, that very strong bounds are also obtained from the analysis of solar neutrino data~\cite{deHolanda:2019tuf}, which can only be applied to state-selection models. 

While in the literature a lot of focus has been set on state-selection and phase-perturbation, in this work we calculate the bounds on neutrino quantum decoherence from the analysis of SN1987A data collected in the Kamiokande-II~\cite{Kamiokande-II:1987idp}, Baksan~\cite{Alekseev:1988gp} and IMB~\cite{Bionta:1987qt} experiments for a decoherence model where neutrino loss is possible. Neutrinos which encounter a virtual black hole in their trajectory might be absorbed and immediately ejected as a different particle in a random direction (remind that only total energy, charge and angular momentum are conserved), thus never reaching the detector. The fact that we observed neutrinos from SN1987A where neutrinos traveled a very long distance allows us to set very strong bounds on the parameters for the neutrino-loss model.

\section{The formalism}
\label{sec:formalism}

We work in the open quantum systems framework. The time-evolution of the neutrinos is described through the Lindblad  master equation~\cite{Lindblad:1975ef,Gorini:1975nb}

\begin{equation}
	\frac{\partial\rho_\nu(t)}{\partial t}=-i[H,\rho_\nu(t)] - \mathcal{D}[\rho_\nu(t)]\,,
    \label{eq:rho_time_dep}
\end{equation}
where $\rho_\nu(t)$ is a Hermitian density matrix describing the neutrino states in the mass basis and $H$ is the Hamiltonian of the neutrino subsystem. Vacuum propagation and standard matter effects are encoded in $H$. The operator $\mathcal{D}[\rho_\nu(t)]$ describes the interaction of the neutrino with the environment and contains therefore all possible decoherence effects. The exact form of this operator depends on the decoherence model of interest.  In general, it is also a Hermitian matrix which can be expressed as~\cite{Gago:2002na,Benatti:2000ph,Lisi:2000zt,Stuttard:2020qfv}

\begin{equation}
\label{eq:DecTerm}
\mathcal{D}[\rho_\nu(t)]=\frac{1}{2}\sum_{j=1}^{N^2-1} \left([\mathcal{O}_j,\rho_\nu(t) \mathcal{O}_j^{\dagger}]+[\mathcal{O}_j\rho_\nu(t), \mathcal{O}_j^{\dagger}] \right) \, ,
\end{equation}
where $N$ is the dimension of the $\mathrm{SU}(N)$ Hilbert space defining the neutrino system (we will assume only the case $N = 3$) and $\mathcal{O}_j$ are dissipative operators (also $N \times N$ complex matrices) which characterize the coupling of the neutrino subsystem to its environment. Let us notice that, while being Hermitian, the matrix $\mathcal{D}[\rho_\nu(t)]$ can lead to nonunitary neutrino propagation due to the interactions of the neutrino subsystem with the environment as is the case when considering possible neutrino loss along the neutrino's trajectory.
We can expand the operators in Eq~\eqref{eq:DecTerm} in terms of the generators from the $\mathrm{SU}(N)$ group~\cite{Benatti:2000ph,Gago:2002na,Carrasco:2018sca,Buoninfante:2020iyr,Stuttard:2020qfv}

\begin{equation}
\label{eq:GellMannexp}
\mathcal{D}[\rho_\nu(t)] = c_k \lambda^k \,,
\end{equation} 
with $\rho_\nu=\sum \rho_\nu^k \lambda^k$ and $\mathcal{O}_j=\sum \mathcal{O}^j_k \lambda^k$. In the previous expression, $c_k$ are the coefficients of the expansion, and the index $k$ runs from 0 to 8 (for $N=3$),  with $\lambda^0$ being the identity matrix and $\lambda^k$ the Gell-Mann matrices satisfying $[\lambda^a,\lambda^b]=i\sum_c f_{abc}\lambda^c$, $f_{abc}$ being the structure constants of $\mathrm{SU}(3)$.
We can then express the decoherence term as $\mathcal{D}[\rho_\nu(t)] = (\mathbf{D}_{k \ell}~ \rho_\nu^\ell) \lambda^k$, where $\rho_\nu^\ell$ are the coefficients of the neutrino density matrix, and $\mathbf{D}_{k \ell} ~\rho_\nu^\ell = c_k$ contains the elements of a $N^2 \times N^2$ matrix representing the free parameters of the system, which in its most general form can be parameterized as

\begin{equation}
\label{eq:Dmatrixfull}
\bf{D}=
\begin{pmatrix}
\Gamma_{0} & \beta_{01} & \beta_{02} & \beta_{03} & \beta_{04} & \beta_{05} & \beta_{06} & \beta_{07} & \beta_{08} \\ 
\beta_{01} & \Gamma_{1} & \beta_{12} & \beta_{13} & \beta_{14} & \beta_{15} & \beta_{16} & \beta_{17} & \beta_{18} \\ 
\beta_{02} & \beta_{12} & \Gamma_{2} & \beta_{23} & \beta_{24} & \beta_{25} & \beta_{26} & \beta_{27} & \beta_{28} \\ 
\beta_{03} & \beta_{13} & \beta_{23} & \Gamma_{3} & \beta_{34} & \beta_{35} & \beta_{36} & \beta_{37} & \beta_{38} \\ 
\beta_{04} & \beta_{14} & \beta_{24} & \beta_{34} & \Gamma_{4} & \beta_{45} & \beta_{46} & \beta_{47} & \beta_{48} \\ 
\beta_{05} & \beta_{15} & \beta_{25} & \beta_{35} & \beta_{45} & \Gamma_{5} & \beta_{56} & \beta_{57} & \beta_{58} \\ 
\beta_{06} & \beta_{16} & \beta_{26} & \beta_{36} & \beta_{46} & \beta_{56} & \Gamma_{6} & \beta_{67} & \beta_{68} \\ 
\beta_{07} & \beta_{17} & \beta_{27} & \beta_{37} & \beta_{47} & \beta_{57} & \beta_{67} & \Gamma_{7} & \beta_{78} \\ 
\beta_{08} & \beta_{18} & \beta_{28} & \beta_{38} & \beta_{48} & \beta_{58} & \beta_{68} & \beta_{78} & \Gamma_{8} \\
\end{pmatrix}\,,
\end{equation}	
where all entries are real scalars. Note that even though $\bf{D}$ is a $9\times9$ matrix, the resulting $\mathcal{D}[\rho_\nu(t)]$ is still a $3\times3$ matrix, which is obvious from the connection $\mathcal{D}[\rho_\nu(t)] = (\mathbf{D}_{k \ell}~ \rho_\nu^\ell) \lambda^k$. Given the large number of free parameters that this matrix contains, an analysis including all parameters is realistically not practical. Often focus is set on the diagonal parameters\footnote{Note, however, that the presence of non-diagonal parameters can lead to very interesting phenomenology, too~\cite{Carrasco:2018sca,Buoninfante:2020iyr}.}. The above mentioned models correspond to~\cite{Stuttard:2020qfv}

\begin{eqnarray}
    \bf{D}_{\textrm{phase-pert.}} &=& \diag(0,\Gamma,\Gamma,0,\Gamma,\Gamma,\Gamma,\Gamma,0)\,,
    \\
    \bf{D}_{\textrm{state-select}} &=& \diag(0,\Gamma,\Gamma,\Gamma,\Gamma,\Gamma,\Gamma,\Gamma,\Gamma)\,,
    \\
    \bf{D}_{\nu-\textrm{loss}} &=& \diag(\Gamma,\Gamma,\Gamma,\Gamma,\Gamma,\Gamma,\Gamma,\Gamma,\Gamma)\,.
\end{eqnarray}
It is assumed that the new parameter $\Gamma$ can have the following energy dependence

\begin{equation}
\label{eq:gamma_E}
\Gamma = \gamma_{0} \left( \frac{E_\nu}{E_0} \right)^n \, ,
\end{equation}
where $E_\nu$ is the neutrino energy and $E_0$ is a reference energy. For easy comparison with existing literature we set $E_0=1$~GeV, even though this energy is far out of the range of the energies of supernova neutrinos. 

Notice that if we consider $n=-1$ the phenomenology of neutrino decoherence is similar to that of neutrino decay, although decoherence affects all neutrinos, while in the case of neutrino decay, each neutrino $\nu_i$ can have a different decay parameter associated with it. The case $n=-2$ can arise naturally when considering a coupling among neutrinos and gravitational waves~\cite{Domi:2024ypm}. If the loss of coherence is related to quantum gravity, one could naturally expect $n$ to be positive, but some models also predict negative values. For some concrete models, see Refs.~\cite{DEsposito:2023psn,Nandi:2024wkm,Domi:2024ypm,Domi:2024muy}. For completeness and comparison with the literature, we consider the cases $n=-2,-1,0,1,2$.

The leading bounds on $\gamma_0$ are obtained in Refs.~\cite{DeRomeri:2023dht,ICECUBE:2023gdv}, depending on the value of $n$. The bounds have been calculated for the phase-perturbation and state-selection models, and the numerical values are very similar. It can be expected that slightly stronger bounds would be obtained for the neutrino-loss model, since, in addition to decoherence effects, we would observe an overall reduction of the neutrino flux at the detector. The state-selection and phase-perturbation models cannot be tested using supernova neutrinos from SN1987A, because they only lead to slightly different flux compositions, not sufficiently different from the standard case to be distinguished. It was shown in Ref.~\cite{dosSantos:2023skk} that even future experiments could place bounds on these models only for a relatively near supernova.

However, if neutrinos are lost along the trajectory, the overall flux of neutrinos at the detector is reduced, and strong bounds can be obtained from supernova data. Note that the same behavior can be observed when comparing the SN1987A bounds on invisible~\cite{Martinez-Mirave:2024hfd} and visible~\cite{Ivanez-Ballesteros:2023lqa} neutrino decay. From here on we will focus only on this case where neutrino-loss occurs. In this scenario, the oscillation probability is simply given by~\cite{dosSantos:2023skk}

\begin{equation}
    P_{\alpha\beta}^{\textrm{dec}}(L,E_\nu) = P_{\alpha\beta}^{\textrm{sm}}e^{-L\gamma_0(E_\nu/E_0)^n}\,.
    \label{eq:dec_prob}
\end{equation}
where $P_{\alpha\beta}^{\textrm{sm}}$ is the standard model oscillation probability and $L$ is the distance traveled by the neutrino. 

\section{Data analysis}
\label{sec:data}
In order to provide very robust bounds we perform two different analyses of SN1987A data. We first consider a time-integrated (TI) analysis following the approach described in Ref.~\cite{Fiorillo:2023frv} and then we perform a time-dependent (TD)  analysis based on the procedure outlined in Ref.~\cite{Pagliaroli:2008ur}. \\


{\em TI analysis -} We consider the time-integrated event rate produced by electron antineutrinos interacting through inverse beta decay. The expected signal $\bar{R}(E_e)$ (differential in the positron energy $E_e$) is given by:
\begin{eqnarray}
    \bar{R}(E_e) &=& N_p \sigma_{\overline{\nu}_e p}(E_\nu)F_{\overline{\nu}_e}(E_\nu)\eta(E_e)\,,
    \label{eq:ratesimp}
\end{eqnarray}
where $E_\nu$ the neutrino energy, $N_p$ is the number of free protons in the detector, $\sigma_{\overline{\nu}_e p}$ is the inverse beta decay (IBD) cross section \cite{Strumia:2003zx}, $\eta$ the detector dependent detection efficiency and $F_{\overline{\nu}_e}$ is the electron antineutrino fluence.
In the above expression, we neglect the (small) kinetic energy of the neutron in the final state and we relate the neutrino and positron energy by $E_{e} = E_\nu - (m_n - m_p)$, where $m_n$ and $m_p$ are neutron and proton masses, respectively. 
The number of target protons $N_p$ is calculated by considering $2.14$ kt ($6.8$ kt) of H$_2$O for KII (IMB) detectors and $280$ tons of water equivalent for the Baksan scintillator.
For the detection efficiencies we adopt the ones reported in Fig.~1 of Ref. ~\cite{Loredo:2001rx}. 
The standard model electron antineutrino fluence, i.e. the time integrated neutrino flux, is assumed to be well described by a Maxwell-Boltzmann spectrum in the form
\begin{eqnarray}
F^{\text{sm}}_{\overline{\nu}_e}(E_\nu)&=&\frac{1}{4\pi D^2}\frac{E^{\overline{\nu}_e}_{\text{tot}}}{\langle E_{\overline{\nu}_e}\rangle}
\frac{(1+\alpha)^{(1+\alpha)}}{\Gamma[1+\alpha]}\nonumber\\
&\times& \left(\frac{E_\nu}{\langle E_{\overline{\nu}_e}\rangle}\right)^2e^{-\frac{(1+\alpha)E_\nu}{\langle E_{\overline{\nu}_e}\rangle}}\,,
\end{eqnarray}
where we set $\alpha=2$ and $D=50$~kpc is the distance to SN1987A.
The two free parameters in this time-independent data analysis are: the total energy $E^{\overline{\nu}_e}_{\text{tot}}$ emitted in $\overline{\nu}_e$ and the average $\overline{\nu}_e$ energy $\langle E_{\overline{\nu}_e}\rangle$.
As discussed in \cite{Fiorillo:2023frv}, this description can fit the time-integrated numerical fluences emerging from a SN, whereas the instantaneous neutrino fluxes can differ from this basic assumption showing pronounced pinching.
Since the total energy and the time-integrated average energies of $\overline{\nu}_e$ and $\overline{\nu}_{x}$ are expected to be nearly the same, oscillations among different flavours (or state selection and phase-perturbation decoherence) do not alter the predicted $\overline{\nu}_e$ fluence at Earth.
The effect of decoherence with neutrino loss can instead become relevant and it is simply described as
\begin{equation}
    F_{\overline{\nu}_e}^{\textrm{dec}}(E_\nu) = F_{\overline{\nu}_e}^{\textrm{sm}}(E_\nu)e^{-D\gamma_0(E_\nu/E_0)^n}\,,
\end{equation}
where we propagated the decoherence term from the probability (see Eq.~\eqref{eq:dec_prob}) to the fluence at Earth.
The individual likelihoods for each detector are given in this case by
\begin{equation}
    \mathcal{\overline{L}}_d = e^{-\overline{R} }\prod_{i=1}^{N^*_d}
    \left[\int\overline{R}(E_e)G(E_e,E_i)dE_e\right]\,,
\end{equation}
where the $\overline{R}$ is the rate in Eq.~\eqref{eq:ratesimp} integrated over the total energy range, while 
\begin{equation}
G(E_e,E_i) = \frac{1}{\sqrt{2\pi}\delta E_i}\exp\left(-\frac{(E_e-E_i)^2}{2\delta E_i^2}\right)   
\end{equation}
are the energy smearing functions with $E_i$ and $\delta E_i$ being the measured positron energies and uncertainties, respectively.
In a TI analysis it is not clear how to treat backgrounds and for the present analysis we have not included them. Following Ref.~\cite{Fiorillo:2023frv} we define a priori the data set of Kamiokande events $N^*_d$ which we analyze excluding the events below the energy threshold. The list of considered events is indicated with a $^*$ in Tab.~\ref{tab:Kam_events}. For the other experiments all events are included and are summarized in Ref.~\cite{Loredo:2001rx}.\\


\begin{table}
    \centering
    \begin{tabular}{|c|c|c|c|}
    \hline
        $\delta t_{i}$ & $E_{i}\pm\delta E_{i}$ & $\theta_i$  & $B_i$ \\
        \hline
        $0.0^*$ & $20.0\pm2.9$  &$18\pm18$  &  $1.0\times10^{-5}$\\
        $0.107^*$ & $13.5\pm3.2$ & $40\pm27$ & $5.4\times10^{-4}$ \\
        $0.303^*$ & $7.5\pm2.0$  & $108\pm32$ &  $3.1\times10^{-2}$ \\
        $0.324^*$ & $9.2\pm2.7$ & $70\pm30$ &  $8.5\times10^{-3}$ \\
        $0.507^*$ & $12.8\pm2.9$ & $135\pm23$  &  $5.3\times10^{-4}$ \\
        0.686 & $6.3\pm1.7$ & $68\pm77$ &  $7.1\times10^{-2}$ \\
         $1.541^*$& $35.4\pm8.0$ & $32\pm16$ &  $5.0\times10^{-6}$ \\
         $1.728^*$& $21.0\pm4.2$ & $30\pm18$  &  $1.0\times10^{-5}$ \\
         $1.915^*$& $19.8\pm3.2$& $38\pm22$  &  $1.0\times10^{-5}$ \\
         $9.219^*$& $8.6\pm2.7$& $122\pm 30$ &  $1.8\times10^{-2}$\\
         $10.433^*$ & $13.0\pm2.6$ & $49\pm26$& $4.0\times10^{-4}$\\
         $12.439^*$ & $8.9\pm1.9$& $91\pm39$& $1.4\times10^{-2}$\\
         $17.641$ &$6.5\pm1.6$& $103\pm50$ & $7.2\times10^{-3}$\\
         $20.257$ & $5.4\pm1.4$ & $110\pm50$ & $5.2\times10^{-2}$\\
         $21.355$ & $4.6\pm1.3$ & $120\pm50$ & $1.8\times10^{-2}$\\
         $23.814$& $6.5\pm1.6$ & $112\pm50$ & $7.3\times10^{-2}$\\
         \hline
    \end{tabular}
    \caption{The Kamiokande-II data set. For the TI data analysis only the data indicated with a $^*$ are considered, while in the TD analysis we include the entire data set. Energies are given in MeV and times are relative times with respect to the bounce in seconds.}
    \label{tab:Kam_events}
\end{table}

{\em TD analysis -}
The signal rate for each detector, triply differential in time $t$, positron energy $E_e$ and cosine of the positron emission angle $c_\theta$,
is computed according to
\begin{eqnarray}
    R(t, E_e, c_\theta) &=& N_p \frac{d\sigma_{\overline{\nu}_e p}}{dc_\theta}(E_\nu,c_\theta)\Phi_{\overline{\nu}_e}(t,E_\nu)\nonumber
    \\
    &\times&\xi(c_\theta)\eta(E_e)\frac{dE_\nu}{dE_e}\,,
    \label{eq:rate}
\end{eqnarray}
where $\xi$ is the angular bias (1 for Kamiokande-II and Baksan and $1+0.1c_\theta$ for IMB), and, $\Phi_{\overline{\nu}_e}$ is the electron antineutrino flux.
In the above equation, the neutrino and positron energies are related via
\begin{equation}
    E_\nu = \frac{E_e + \delta_{-}}{1-(E_e-p_ec_\theta)/m_p}\,,
\end{equation}
where $p_e$ is the positron momentum and $\delta_-=(m_n^2-m_p^2-m_e^2)/(2m_p)=1.294$~MeV, with $m_p$, $m_n$, and $m_e$ being the masses of the proton, neutron and electron, respectively.
Regarding the differential IBD cross section we use the approximated formulas valid for the energies under consideration presented in Refs.~\cite{Strumia:2003zx,Vissani:2014doa}.

The electron antineutrino flux is parameterized as a pure cooling emission phase, i.e. a thermal emission from the neutrinosphere: 
\begin{equation}
   \Phi_{\overline{\nu}_e}^0(t,E_\nu) = \frac{R_c^2g_{\overline{\nu}_e}(t,T_c(t))}{8\pi^2 D^2} 
\end{equation}
with the Fermi-Dirac spectrum 
\begin{equation}
    g_{\overline{\nu}_e}(t,T_c(t)) = \frac{E_\nu^2}{1+\exp(E_\nu/T_c(t))},
\end{equation}
which depends on 
\begin{equation}
    T_c(t) = T_c \exp(-t/(4\tau_c))\,.
\end{equation}
The physical parameters in this scenario are the radius of the neutrinosphere $R_c$, the initial temperature $T_c$ and the time constant of the process $\tau_c$. 
The initial flux of muon and tau antineutrinos $\Phi_{\overline{\nu}_x}^0(t,E_\nu)$ is the same as for electron antineutrinos, but with the replacements $T_c\rightarrow 1.2 T_c$ and $R_c\rightarrow R_c/1.2^2$.
This prescription implements equipartion of emitted energy among different flavours. However, it takes into account that, due to reduced interactions, muon and tau neutrinos have a more internal neutrinosphere and a slightly larger average energy than electron neutrinos.
Since emission spectra of $\overline{\nu}_e$ and $\overline{\nu}_x$ are not identical, we need to account for possible conversions among neutrino flavours.
In the standard oscillation scenario the flux at the detector is given by 
\begin{equation}
    \Phi_{\overline{\nu}_e}^{\textrm{sm}}(t,E_\nu) = P_{ee} \Phi_{\overline{\nu}_e}^0(t,E_\nu) + P_{xe} \Phi_{\overline{\nu}_x}^0(t,E_\nu).
\end{equation}
%
The neutrino oscillation probabilities are given by $P_{ee} = \cos^2\theta_{12}\cos^2\theta_{13} \sim 0.7$ and $P_{xe}=1-P_{ee}\sim 0.3$, where we use~\cite{deSalas:2020pgw} $\sin^2\theta_{12}=0.318$ and $\sin^2\theta_{13} = 0.022$.
Even though the neutrino mass ordering is not definitely determined yet~\cite{Gariazzo:2022ahe}, we consider here only normal mass ordering, slightly preferred by the data~\cite{deSalas:2020pgw,Gariazzo:2022ahe,Capozzi:2021fjo,Esteban:2024eli}.
Phase-perturbation decoherence models do not alter this prediction because they have the same effects of kinematic decoherence that is already taken into account for supernova neutrinos.
State selection decoherence models lead to equipartition of flavours, corresponding to $P_{ee} = 1/3$ in the above expression.
The different flux composition however, due to similarities among $\Phi_{\overline{\nu}_e}^0$ and $\Phi_{\overline{\nu}_X}^0$, cannot be constrained by SN1987A data. 
It was shown in Ref.~\cite{dosSantos:2023skk} that even future experiments could place bounds on these models only for a relatively near supernova.
Decoherence models with neutrino loss can instead have observable effects, since the flux at the detector becomes:
\begin{equation}
    \Phi_{\overline{\nu}_e}^{\textrm{dec}}(t,E_\nu) = \Phi_{\overline{\nu}_e}^{\textrm{sm}}(t,E_\nu)e^{-D\gamma_0(E_\nu/E_0)^n}\,,
\end{equation}
where we again propagated the decoherence term from the probability, Eq.~\eqref{eq:dec_prob}, to the flux at Earth. 


The statistical analysis is performed by evaluating 

\begin{equation}
    \chi^2 = -2 \sum_{d = K,I,B} \log\mathcal{L}_d
\end{equation}
where the individual likelihoods for each detector are given by

\begin{eqnarray}
    \mathcal{L}_d &=& e^{-f_d\int R(t)dt}\prod_{i=1}^{N_d}e^{R(t_i)\tau_d}\nonumber
    \\
    &\times&\left[\frac{B_i}{2} + \int R(t_i,E_e,c_{\theta_i})G(E_e,E_i)dE_e\right]\,,
\end{eqnarray}
where the product is taken over the $N_d$ events observed in each detector, $R(t)$ is Eq.~\eqref{eq:rate} integrated over $E_e$ and $c_\theta$, $B_i$ are the background rates for each detector. 
 The quantities with subscript $i$ are the measured times ($t_i$), energies ($E_i\pm\delta E_i$), and angles ($\theta_i$) of the $i$th event. Finally, $f_d$ and $\tau_d$ are the live-time fraction and dead-time of the detectors. We take $f_d = 0.9055$ ($f_d = 1$) and $\tau_d = 0.035$~s ($\tau_d = 0$~s) for IMB (Kamiokande-II and Baksan).

\section{Results and discussion}
\label{sec:results}

We now discuss the results of our analyses.
Note that the bounds on the decoherence parameter $\gamma_0$ are obtained by marginalizing over the parameters used to describe SN1987A neutrino emission.
These parameters are based on different physical considerations in the two analyses performed in this work.
Specifically,  the total energy and the average energy of emitted electron antineutrinos, $E^{\overline{\nu}_e}_{\text{tot}}$ and $\langle E_\nu\rangle$, are considered in the TI analysis, while the radius and temperature of neutrinosphere, $R_c$ and $T_c$, and the time scale $\tau_c$ for  supernova cooling are adopted in the TD analysis.
These parameters  are varied in very conservative ranges.
For the TI analysis the total energy emitted in $\overline{\nu}_e$ is bounded to be smaller than the neutron star binding energy, i.e. $E^{\overline{\nu}_e}_{\text{tot}}\leq5\cdot10^{53}$ ergs, while the average energy of $\overline{\nu}_e$ is constrained to be $\langle E_\nu\rangle<90$ MeV.
Note that the upper limit for $E^{\overline{\nu}_e}_{\text{tot}}$ corresponds to the extreme assumption that all available energy is radiated in electron antineutrinos.
For the TD analysis, we assume the following priors,  $R_c<100$~km, $2~\textrm{MeV}<T_c<8~\textrm{MeV}$, and $1~\textrm{s}<\tau_c<10~\textrm{s}$, based on general expectations from supernova theory and recent simulations \cite{Janka:2025tvf,Fiorillo:2023frv,Garching}.

\begin{figure}[!t]
\centering
\includegraphics[width=0.49\textwidth]{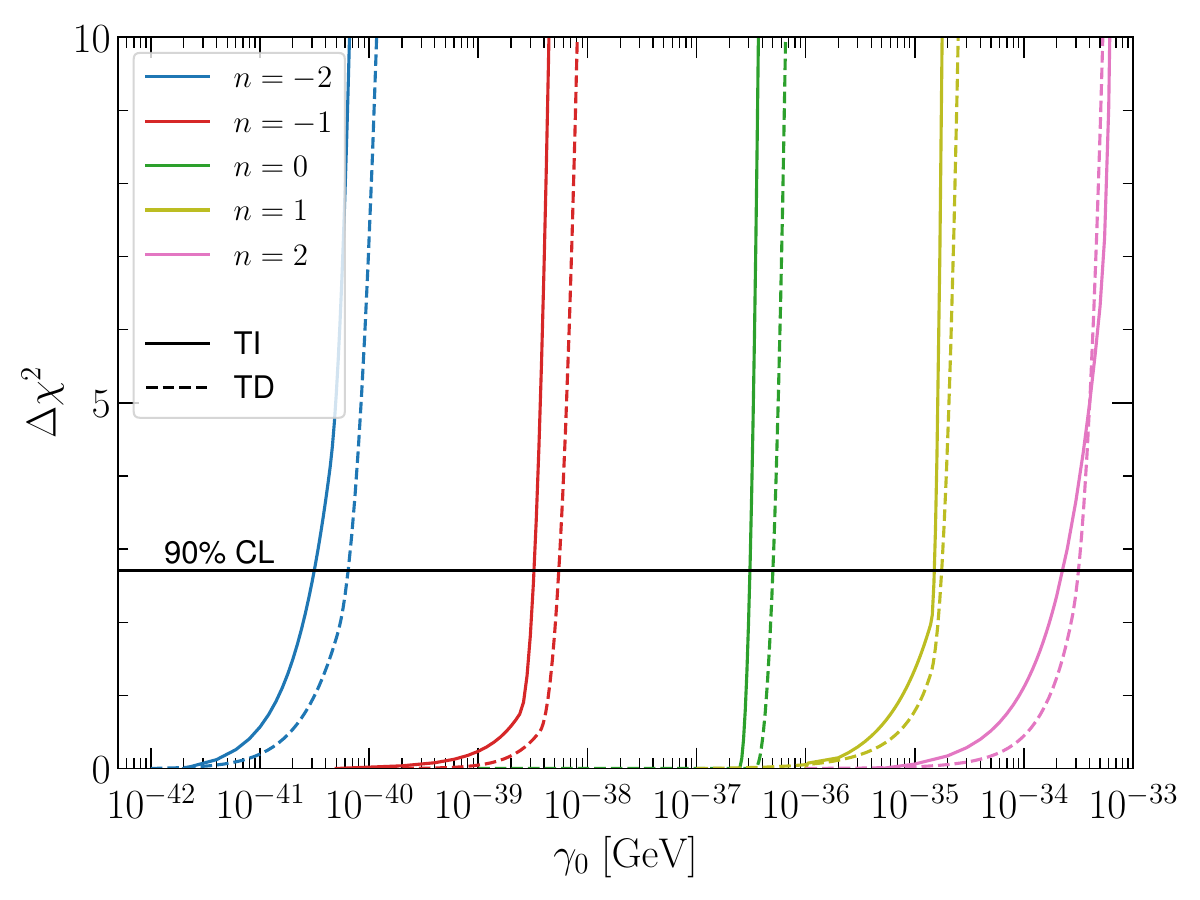}
\caption{The $\chi^2$-profiles for the decoherence parameter $\gamma_0$ for different values of $n$ as indicated in the legend. The solid (dashed) lines correspond to the TI (TD) analyses.}
\label{fig:deltachi2}
\end{figure}

The 90\% confidence level (CL) bounds on the decoherence parameter $\gamma_0$ obtained with the time-integrated (TI) and the time-dependent (TD) approach are shown in Fig.~\ref{fig:deltachi2} and Tab.~\ref{tab:bounds}.
%
Since our reference energy in Eq.~\eqref{eq:gamma_E} is $E_0 = 1$~GeV while the average energy of supernova neutrinos is $\sim 10$~MeV, we find the strongest bound on $\gamma_0$ for $n=-2$, becoming then weaker as $n$ increases. 
As becomes evident from the table, the bound from SN1987A is for all $n$ many orders of magnitude larger than the previous bounds for which we only indicate an order of magnitude estimate, since Refs.~\cite{deHolanda:2019tuf, ICECUBE:2023gdv, DeRomeri:2023dht} did not consider neutrino-loss models. 

\begin{table}[t]
    \centering
    \begin{tabular}{|c|c|c|c|}
\hline
         $n$ & previous bound & this work (TI) & this work (TD)  \\\hline
         ~~-2~~ & $\mathcal{O}(10^{-27})$ & $3.1\times10^{-41}$ & $6.4\times10^{-41}$ \\
         ~~-1~~ & $\mathcal{O}(10^{-24})$ & $3.2\times10^{-39}$ & $5.5\times10^{-39}$ \\
         ~~0~~ & $\mathcal{O}(10^{-28})$ & $3.1\times10^{-37}$ & $5.0\times10^{-37}$ \\
         ~~1~~ & $\mathcal{O}(10^{-28})$ & $1.5\times10^{-35}$ & $1.8\times10^{-35}$ \\
         ~~2~~ & $\mathcal{O}(10^{-32})$ & $2.3\times10^{-34}$ & $3.2\times10^{-34}$ \\
         \hline
    \end{tabular}
    \caption{\label{tab:bounds} The bounds on the decoherence parameter $\gamma_0$ in GeV at 90\% CL obtained in this paper with the time-integrated (TI) and the time-dependent (TD) approach. For comparison we also show an order of magnitude estimate from the leading bounds from other experiments, Ref.~\cite{DeRomeri:2023dht} for $n<0$, Ref.~\cite{deHolanda:2019tuf} for $n=0$ and Ref.~\cite{ICECUBE:2023gdv} for $n>0$.}
\end{table}

The agreement between the results obtained by using different approaches is striking and shows us that the adopted parameterization for the supernova emission is not relevant for the final results.
%
%
Indeed, the TI approach is mainly based on the spectral shape of the supernova signal observed in the numerical simulation without taking into account the temporal evolution of the emission and the temporal distribution of the observed events. 
Since the neutrino average energy can be constrained by the observed signal, the total number of observed events translate into a constraint on $\gamma_0$, once an upper limit on the total energy emitted in electron antineutrinos is given. This, however, can be conservatively estimated as $E^{\overline{\nu}_e}_{\text{tot}}\leq5\cdot10^{53}$ ergs, as it was explained above.
The number of free parameters in this analysis is smaller (only $3$); however, background rejection is not possible, requiring that a reduced data-set (where background events are expected to be absent) is defined "a priori".
%
%
On the other hand, the TD analysis is more complex. The signal is parameterized according to a phenomenological description of the neutrino emission during the cooling phase and the number of free parameters is larger ($4$).
However, the time (and angular) distribution of the expected supernova neutrino events can be used to separate signal and background contributions. 
In this case, considering that the energy and time distribution of observed events can be used to constrain $T_{c}$ and $\tau_{c}$ parameters, the bound on $\gamma_0$ is connected with the assumed radius of the neutrinosphere, for which the very conservative upper bound $R_c < 100$~km is adopted.
We checked that the best-fit obtained in absence of decoherence are in perfect agreement with the ones obtained in previous analysis of SN1987A, i.e Ref.~\cite{Fiorillo:2023frv} for the TI approach and Ref.~\cite{Pagliaroli:2008ur} for the TD analysis.  

Finally, we comment on the specific case of $n=2$: the $\Gamma$ parameter can be expressed relative to the Planck scale as $\Gamma=\xi_{\rm{Planck}}E^n/M_{\rm{Planck}}^{n-1}$ with $M_{\rm{Planck}}\simeq1.2\times 10^{19}$ GeV and $\xi_{\rm{Planck}}$ being a relative strength for $\nu$-virtual black hole interactions, see Ref.~\cite{Stuttard:2020qfv}. Our bound on $\gamma_0$ can be converted into a limit on this strength, giving $\xi_{\rm{Planck}}<2.8\times 10^{-15}$ (to be compared with the ones reported in \cite{ICECUBE:2023gdv} from IceCube of $\xi_{\rm{Planck}}<1.2\times 10^{-13}$).

Note that for even larger values of $n$ than those considered here, the bound from Ref.~\cite{ICECUBE:2023gdv} becomes stronger than the bound from SN1987A, since in these cases the shorter distance of atmospheric neutrinos becomes compensated for by their much larger energies. In principle, the strongest bounds could be obtained by looking at extremely high energy events from far away sources. However, since sources are not easily identified (or modeled), we chose to discuss SN1987A here, for which we have a clear theoretical description. We will discuss the case of high-energy neutrinos in a future work.


%
\section*{Acknowledgements}
 The work of GP and FLV is partially supported by  grant number 2022E2J4RK "PANTHEON: Perspectives in Astroparticle and
Neutrino THEory with Old and New messengers" under the program PRIN 2022 funded by the Italian Ministero dell’Universit\`a e della Ricerca (MUR) and by the European Union – Next Generation EU.

\bibliographystyle{utphys}
\bibliography{bibliography}

\providecommand{\href}[2]{#2}\begingroup\raggedright\begin{thebibliography}{10}

\bibitem{Giunti:1991sx}
C.~Giunti, C.~W. Kim, and U.~W. Lee, ``{Coherence of neutrino oscillations in
  vacuum and matter in the wave packet treatment},''
  \href{http://dx.doi.org/10.1016/0370-2693(92)90308-Q}{{\em Phys. Lett. B}
  {\bfseries 274} (1992) 87--94}.

\bibitem{Kiers:1995zj}
K.~Kiers, S.~Nussinov, and N.~Weiss, ``{Coherence effects in neutrino
  oscillations},'' \href{http://dx.doi.org/10.1103/PhysRevD.53.537}{{\em Phys.
  Rev. D} {\bfseries 53} (1996) 537--547},
  \href{http://arxiv.org/abs/hep-ph/9506271}{{\ttfamily arXiv:hep-ph/9506271}}.

\bibitem{Ohlsson:2000mj}
T.~Ohlsson, ``{Equivalence between neutrino oscillations and neutrino
  decoherence},'' \href{http://dx.doi.org/10.1016/S0370-2693(01)00178-2}{{\em
  Phys. Lett. B} {\bfseries 502} (2001) 159--166},
  \href{http://arxiv.org/abs/hep-ph/0012272}{{\ttfamily arXiv:hep-ph/0012272}}.

\bibitem{Beuthe:2002ej}
M.~Beuthe, ``{Towards a unique formula for neutrino oscillations in vacuum},''
  \href{http://dx.doi.org/10.1103/PhysRevD.66.013003}{{\em Phys. Rev. D}
  {\bfseries 66} (2002) 013003},
  \href{http://arxiv.org/abs/hep-ph/0202068}{{\ttfamily arXiv:hep-ph/0202068}}.

\bibitem{Giunti:2003ax}
C.~Giunti, ``{Coherence and wave packets in neutrino oscillations},''
  \href{http://dx.doi.org/10.1023/B:FOPL.0000019651.53280.31}{{\em Found. Phys.
  Lett.} {\bfseries 17} (2004) 103--124},
  \href{http://arxiv.org/abs/hep-ph/0302026}{{\ttfamily arXiv:hep-ph/0302026}}.

\bibitem{Blennow:2005yk}
M.~Blennow, T.~Ohlsson, and W.~Winter, ``{Damping signatures in future neutrino
  oscillation experiments},''
  \href{http://dx.doi.org/10.1088/1126-6708/2005/06/049}{{\em JHEP} {\bfseries
  06} (2005) 049}, \href{http://arxiv.org/abs/hep-ph/0502147}{{\ttfamily
  arXiv:hep-ph/0502147}}.

\bibitem{Farzan:2008eg}
Y.~Farzan and A.~Y. Smirnov, ``{Coherence and oscillations of cosmic
  neutrinos},'' \href{http://dx.doi.org/10.1016/j.nuclphysb.2008.07.028}{{\em
  Nucl. Phys. B} {\bfseries 805} (2008) 356--376},
  \href{http://arxiv.org/abs/0803.0495}{{\ttfamily arXiv:0803.0495 [hep-ph]}}.

\bibitem{Kayser:2010pr}
B.~Kayser and J.~Kopp, ``{Testing the Wave Packet Approach to Neutrino
  Oscillations in Future Experiments},''
  \href{http://arxiv.org/abs/1005.4081}{{\ttfamily arXiv:1005.4081 [hep-ph]}}.

\bibitem{Naumov:2013uia}
D.~V. Naumov, ``{On the Theory of Wave Packets},''
  \href{http://dx.doi.org/10.1134/S1547477113070145}{{\em Phys. Part. Nucl.
  Lett.} {\bfseries 10} (2013) 642--650},
  \href{http://arxiv.org/abs/1309.1717}{{\ttfamily arXiv:1309.1717
  [quant-ph]}}.

\bibitem{Jones:2014sfa}
B.~J.~P. Jones, ``{Dynamical pion collapse and the coherence of conventional
  neutrino beams},'' \href{http://dx.doi.org/10.1103/PhysRevD.91.053002}{{\em
  Phys. Rev. D} {\bfseries 91} no.~5, (2015) 053002},
  \href{http://arxiv.org/abs/1412.2264}{{\ttfamily arXiv:1412.2264 [hep-ph]}}.

\bibitem{Akhmedov:2019iyt}
E.~Akhmedov, ``{Quantum mechanics aspects and subtleties of neutrino
  oscillations},'' in {\em {International Conference on History of the
  Neutrino}: {1930-2018}}.
\newblock 1, 2019.
\newblock \href{http://arxiv.org/abs/1901.05232}{{\ttfamily arXiv:1901.05232
  [hep-ph]}}.

\bibitem{Grimus:2019hlq}
W.~Grimus, ``{Revisiting the quantum field theory of neutrino oscillations in
  vacuum},'' \href{http://dx.doi.org/10.1088/1361-6471/ab716f}{{\em J. Phys. G}
  {\bfseries 47} no.~8, (2020) 085004},
  \href{http://arxiv.org/abs/1910.13776}{{\ttfamily arXiv:1910.13776
  [hep-ph]}}.

\bibitem{Naumov:2020yyv}
D.~V. Naumov and V.~A. Naumov, ``{Quantum Field Theory of Neutrino
  Oscillations},'' \href{http://dx.doi.org/10.1134/S1063779620010050}{{\em
  Phys. Part. Nucl.} {\bfseries 51} no.~1, (2020) 1--106}.

\bibitem{deGouvea:2020hfl}
A.~de~Gouvêa, V.~De~Romeri, and C.~A. Ternes, ``{Probing neutrino quantum
  decoherence at reactor experiments},''
  \href{http://dx.doi.org/10.1007/JHEP08(2020)049}{{\em JHEP} {\bfseries 08}
  (2020) 018}, \href{http://arxiv.org/abs/2005.03022}{{\ttfamily
  arXiv:2005.03022 [hep-ph]}}.

\bibitem{deGouvea:2021uvg}
A.~de~Gouv\^ea, V.~De~Romeri, and C.~A. Ternes, ``{Combined analysis of
  neutrino decoherence at reactor experiments},''
  \href{http://dx.doi.org/10.1007/JHEP06(2021)042}{{\em JHEP} {\bfseries 06}
  (2021) 042}, \href{http://arxiv.org/abs/2104.05806}{{\ttfamily
  arXiv:2104.05806 [hep-ph]}}.

\bibitem{Akhmedov:2022bjs}
E.~Akhmedov and A.~Y. Smirnov, ``{Damping of neutrino oscillations, decoherence
  and the lengths of neutrino wave packets},''
  \href{http://dx.doi.org/10.1007/JHEP11(2022)082}{{\em JHEP} {\bfseries 11}
  (2022) 082}, \href{http://arxiv.org/abs/2208.03736}{{\ttfamily
  arXiv:2208.03736 [hep-ph]}}.

\bibitem{Krueger:2023skk}
R.~Krueger and T.~Schwetz, ``{Decoherence effects in reactor and Gallium
  neutrino oscillation experiments -- a QFT approach},''
  \href{http://arxiv.org/abs/2303.15524}{{\ttfamily arXiv:2303.15524
  [hep-ph]}}.

\bibitem{deGouvea:2024syg}
A.~de~Gouv\^ea, V.~De~Romeri, and C.~A. Ternes, ``{Addendum to: Combined
  analysis of neutrino decoherence at reactor experiments},''
  \href{http://dx.doi.org/10.1007/JHEP11(2024)095}{{\em JHEP} {\bfseries 11}
  (2024) 095}, \href{http://arxiv.org/abs/2410.01357}{{\ttfamily
  arXiv:2410.01357 [hep-ph]}}.

\bibitem{Gago:2000qc}
A.~M. Gago, E.~M. Santos, W.~J.~C. Teves, and R.~Zukanovich~Funchal, ``{Quantum
  dissipative effects and neutrinos: Current constraints and future
  perspectives},'' \href{http://dx.doi.org/10.1103/PhysRevD.63.073001}{{\em
  Phys. Rev. D} {\bfseries 63} (2001) 073001},
  \href{http://arxiv.org/abs/hep-ph/0009222}{{\ttfamily arXiv:hep-ph/0009222}}.

\bibitem{Benatti:2000ph}
F.~Benatti and R.~Floreanini, ``{Open system approach to neutrino
  oscillations},'' \href{http://dx.doi.org/10.1088/1126-6708/2000/02/032}{{\em
  JHEP} {\bfseries 02} (2000) 032},
  \href{http://arxiv.org/abs/hep-ph/0002221}{{\ttfamily arXiv:hep-ph/0002221}}.

\bibitem{Lisi:2000zt}
E.~Lisi, A.~Marrone, and D.~Montanino, ``{Probing possible decoherence effects
  in atmospheric neutrino oscillations},''
  \href{http://dx.doi.org/10.1103/PhysRevLett.85.1166}{{\em Phys. Rev. Lett.}
  {\bfseries 85} (2000) 1166--1169},
  \href{http://arxiv.org/abs/hep-ph/0002053}{{\ttfamily arXiv:hep-ph/0002053}}.

\bibitem{Benatti:2001fa}
F.~Benatti and R.~Floreanini, ``{Massless neutrino oscillations},''
  \href{http://dx.doi.org/10.1103/PhysRevD.64.085015}{{\em Phys. Rev. D}
  {\bfseries 64} (2001) 085015},
  \href{http://arxiv.org/abs/hep-ph/0105303}{{\ttfamily arXiv:hep-ph/0105303}}.

\bibitem{Breuer:2002pc}
H.~P. Breuer and F.~Petruccione, {\em {The theory of open quantum systems}}.
\newblock 2002.

\bibitem{Morgan:2004vv}
D.~Morgan, E.~Winstanley, J.~Brunner, and L.~F. Thompson, ``{Probing quantum
  decoherence in atmospheric neutrino oscillations with a neutrino
  telescope},''
  \href{http://dx.doi.org/10.1016/j.astropartphys.2006.03.001}{{\em Astropart.
  Phys.} {\bfseries 25} (2006) 311--327},
  \href{http://arxiv.org/abs/astro-ph/0412618}{{\ttfamily
  arXiv:astro-ph/0412618}}.

\bibitem{Anchordoqui:2005gj}
L.~A. Anchordoqui, H.~Goldberg, M.~C. Gonzalez-Garcia, F.~Halzen, D.~Hooper,
  S.~Sarkar, and T.~J. Weiler, ``{Probing Planck scale physics with IceCube},''
  \href{http://dx.doi.org/10.1103/PhysRevD.72.065019}{{\em Phys. Rev. D}
  {\bfseries 72} (2005) 065019},
  \href{http://arxiv.org/abs/hep-ph/0506168}{{\ttfamily arXiv:hep-ph/0506168}}.

\bibitem{Fogli:2007tx}
G.~L. Fogli, E.~Lisi, A.~Marrone, D.~Montanino, and A.~Palazzo, ``{Probing
  non-standard decoherence effects with solar and KamLAND neutrinos},''
  \href{http://dx.doi.org/10.1103/PhysRevD.76.033006}{{\em Phys. Rev. D}
  {\bfseries 76} (2007) 033006},
  \href{http://arxiv.org/abs/0704.2568}{{\ttfamily arXiv:0704.2568 [hep-ph]}}.

\bibitem{Farzan:2008zv}
Y.~Farzan, T.~Schwetz, and A.~Y. Smirnov, ``{Reconciling results of LSND,
  MiniBooNE and other experiments with soft decoherence},''
  \href{http://dx.doi.org/10.1088/1126-6708/2008/07/067}{{\em JHEP} {\bfseries
  07} (2008) 067}, \href{http://arxiv.org/abs/0805.2098}{{\ttfamily
  arXiv:0805.2098 [hep-ph]}}.

\bibitem{Oliveira:2013nua}
R.~L.~N. Oliveira and M.~M. Guzzo, ``{Dissipation and $\theta_{13}$ in neutrino
  oscillations},'' \href{http://dx.doi.org/10.1140/epjc/s10052-013-2434-6}{{\em
  Eur. Phys. J. C} {\bfseries 73} (2013) 2434}.

\bibitem{Oliveira:2016asf}
R.~L.~N. Oliveira, ``{Dissipative Effect in Long Baseline Neutrino
  Experiments},'' \href{http://dx.doi.org/10.1140/epjc/s10052-016-4253-z}{{\em
  Eur. Phys. J. C} {\bfseries 76} no.~7, (2016) 417},
  \href{http://arxiv.org/abs/1603.08065}{{\ttfamily arXiv:1603.08065
  [hep-ph]}}.

\bibitem{BalieiroGomes:2016ykp}
G.~Balieiro~Gomes, M.~M. Guzzo, P.~C. de~Holanda, and R.~L.~N. Oliveira,
  ``{Parameter Limits for Neutrino Oscillation with Decoherence in KamLAND},''
  \href{http://dx.doi.org/10.1103/PhysRevD.95.113005}{{\em Phys. Rev. D}
  {\bfseries 95} no.~11, (2017) 113005},
  \href{http://arxiv.org/abs/1603.04126}{{\ttfamily arXiv:1603.04126
  [hep-ph]}}.

\bibitem{Coelho:2017zes}
J.~A.~B. Coelho, W.~A. Mann, and S.~S. Bashar, ``{Nonmaximal $\theta_{23}$
  mixing at NOvA from neutrino decoherence},''
  \href{http://dx.doi.org/10.1103/PhysRevLett.118.221801}{{\em Phys. Rev.
  Lett.} {\bfseries 118} no.~22, (2017) 221801},
  \href{http://arxiv.org/abs/1702.04738}{{\ttfamily arXiv:1702.04738
  [hep-ph]}}.

\bibitem{Coelho:2017byq}
J.~A.~B. Coelho and W.~A. Mann, ``{Decoherence, matter effect, and neutrino
  hierarchy signature in long baseline experiments},''
  \href{http://dx.doi.org/10.1103/PhysRevD.96.093009}{{\em Phys. Rev. D}
  {\bfseries 96} no.~9, (2017) 093009},
  \href{http://arxiv.org/abs/1708.05495}{{\ttfamily arXiv:1708.05495
  [hep-ph]}}.

\bibitem{Carpio:2017nui}
J.~Carpio, E.~Massoni, and A.~M. Gago, ``{Revisiting quantum decoherence for
  neutrino oscillations in matter with constant density},''
  \href{http://dx.doi.org/10.1103/PhysRevD.97.115017}{{\em Phys. Rev. D}
  {\bfseries 97} no.~11, (2018) 115017},
  \href{http://arxiv.org/abs/1711.03680}{{\ttfamily arXiv:1711.03680
  [hep-ph]}}.

\bibitem{Coloma:2018idr}
P.~Coloma, J.~Lopez-Pavon, I.~Martinez-Soler, and H.~Nunokawa, ``{Decoherence
  in Neutrino Propagation Through Matter, and Bounds from IceCube/DeepCore},''
  \href{http://dx.doi.org/10.1140/epjc/s10052-018-6092-6}{{\em Eur. Phys. J. C}
  {\bfseries 78} no.~8, (2018) 614},
  \href{http://arxiv.org/abs/1803.04438}{{\ttfamily arXiv:1803.04438
  [hep-ph]}}.

\bibitem{Carpio:2018gum}
J.~A. Carpio, E.~Massoni, and A.~M. Gago, ``{Testing quantum decoherence at
  DUNE},'' \href{http://dx.doi.org/10.1103/PhysRevD.100.015035}{{\em Phys. Rev.
  D} {\bfseries 100} no.~1, (2019) 015035},
  \href{http://arxiv.org/abs/1811.07923}{{\ttfamily arXiv:1811.07923
  [hep-ph]}}.

\bibitem{Carrasco:2018sca}
J.~C. Carrasco, F.~N. D\'\i{}az, and A.~M. Gago, ``{Probing CPT breaking
  induced by quantum decoherence at DUNE},''
  \href{http://dx.doi.org/10.1103/PhysRevD.99.075022}{{\em Phys. Rev. D}
  {\bfseries 99} no.~7, (2019) 075022},
  \href{http://arxiv.org/abs/1811.04982}{{\ttfamily arXiv:1811.04982
  [hep-ph]}}.

\bibitem{Buoninfante:2020iyr}
L.~Buoninfante, A.~Capolupo, S.~M. Giampaolo, and G.~Lambiase, ``{Revealing
  neutrino nature and $CPT$ violation with decoherence effects},''
  \href{http://dx.doi.org/10.1140/epjc/s10052-020-08549-9}{{\em Eur. Phys. J.
  C} {\bfseries 80} no.~11, (2020) 1009},
  \href{http://arxiv.org/abs/2001.07580}{{\ttfamily arXiv:2001.07580
  [hep-ph]}}.

\bibitem{Gomes:2020muc}
A.~L.~G. Gomes, R.~A. Gomes, and O.~L.~G. Peres, ``{Quantum decoherence and
  relaxation in neutrinos using long-baseline data},''
  \href{http://arxiv.org/abs/2001.09250}{{\ttfamily arXiv:2001.09250
  [hep-ph]}}.

\bibitem{Ohlsson:2020gxx}
T.~Ohlsson and S.~Zhou, ``{Density Matrix Formalism for PT-Symmetric
  Non-Hermitian Hamiltonians with the Lindblad Equation},''
  \href{http://dx.doi.org/10.1103/PhysRevA.103.022218}{{\em Phys. Rev. A}
  {\bfseries 103} no.~2, (2021) 022218},
  \href{http://arxiv.org/abs/2006.02445}{{\ttfamily arXiv:2006.02445
  [quant-ph]}}.

\bibitem{Stuttard:2020qfv}
T.~Stuttard and M.~Jensen, ``{Neutrino decoherence from quantum gravitational
  stochastic perturbations},''
  \href{http://dx.doi.org/10.1103/PhysRevD.102.115003}{{\em Phys. Rev. D}
  {\bfseries 102} no.~11, (2020) 115003},
  \href{http://arxiv.org/abs/2007.00068}{{\ttfamily arXiv:2007.00068
  [hep-ph]}}.

\bibitem{Stuttard:2021uyw}
T.~Stuttard, ``{Neutrino signals of lightcone fluctuations resulting from
  fluctuating spacetime},''
  \href{http://dx.doi.org/10.1103/PhysRevD.104.056007}{{\em Phys. Rev. D}
  {\bfseries 104} no.~5, (2021) 056007},
  \href{http://arxiv.org/abs/2103.15313}{{\ttfamily arXiv:2103.15313
  [hep-ph]}}.

\bibitem{Banerjee:2022slh}
I.~K. Banerjee and U.~K. Dey, ``{Neutrino decoherence from generalised
  uncertainty},'' \href{http://dx.doi.org/10.1140/epjc/s10052-023-11565-0}{{\em
  Eur. Phys. J. C} {\bfseries 83} no.~5, (2023) 428},
  \href{http://arxiv.org/abs/2208.12062}{{\ttfamily arXiv:2208.12062
  [hep-ph]}}.

\bibitem{Barenboim:2024zfi}
G.~Barenboim and A.~M. Gago, ``{Quantum decoherence effects: A complete
  treatment},'' \href{http://dx.doi.org/10.1103/PhysRevD.110.095005}{{\em Phys.
  Rev. D} {\bfseries 110} no.~9, (2024) 095005},
  \href{http://arxiv.org/abs/2402.03438}{{\ttfamily arXiv:2402.03438
  [hep-ph]}}.

\bibitem{Hawking:1976ra}
S.~W. Hawking, ``{Breakdown of Predictability in Gravitational Collapse},''
  \href{http://dx.doi.org/10.1103/PhysRevD.14.2460}{{\em Phys. Rev. D}
  {\bfseries 14} (1976) 2460--2473}.

\bibitem{Ellis:1983jz}
J.~R. Ellis, J.~S. Hagelin, D.~V. Nanopoulos, and M.~Srednicki, ``{Search for
  Violations of Quantum Mechanics},''
  \href{http://dx.doi.org/10.1016/0550-3213(84)90053-1}{{\em Nucl. Phys. B}
  {\bfseries 241} (1984) 381}.

\bibitem{Giddings:1988cx}
S.~B. Giddings and A.~Strominger, ``{Loss of Incoherence and Determination of
  Coupling Constants in Quantum Gravity},''
  \href{http://dx.doi.org/10.1016/0550-3213(88)90109-5}{{\em Nucl. Phys. B}
  {\bfseries 307} (1988) 854--866}.

\bibitem{Addazi:2021xuf}
A.~Addazi {\em et~al.}, ``{Quantum gravity phenomenology at the dawn of the
  multi-messenger era\textemdash{}A review},''
  \href{http://dx.doi.org/10.1016/j.ppnp.2022.103948}{{\em Prog. Part. Nucl.
  Phys.} {\bfseries 125} (2022) 103948},
  \href{http://arxiv.org/abs/2111.05659}{{\ttfamily arXiv:2111.05659
  [hep-ph]}}.

\bibitem{deOliveira:2013dia}
R.~L.~N. de~Oliveira, M.~M. Guzzo, and P.~C. de~Holanda, ``{Quantum Dissipation
  and $C\!P$ Violation in MINOS},''
  \href{http://dx.doi.org/10.1103/PhysRevD.89.053002}{{\em Phys. Rev. D}
  {\bfseries 89} no.~5, (2014) 053002},
  \href{http://arxiv.org/abs/1401.0033}{{\ttfamily arXiv:1401.0033 [hep-ph]}}.

\bibitem{BalieiroGomes:2018gtd}
G.~Balieiro~Gomes, D.~V. Forero, M.~M. Guzzo, P.~C. De~Holanda, and R.~L.~N.
  Oliveira, ``{Quantum Decoherence Effects in Neutrino Oscillations at DUNE},''
  \href{http://dx.doi.org/10.1103/PhysRevD.100.055023}{{\em Phys. Rev. D}
  {\bfseries 100} no.~5, (2019) 055023},
  \href{http://arxiv.org/abs/1805.09818}{{\ttfamily arXiv:1805.09818
  [hep-ph]}}.

\bibitem{deHolanda:2019tuf}
P.~C. de~Holanda, ``{Solar Neutrino Limits on Decoherence},''
  \href{http://dx.doi.org/10.1088/1475-7516/2020/03/012}{{\em JCAP} {\bfseries
  03} (2020) 012}, \href{http://arxiv.org/abs/1909.09504}{{\ttfamily
  arXiv:1909.09504 [hep-ph]}}.

\bibitem{JUNO:2021ydg}
{\bfseries JUNO} Collaboration, J.~Wang {\em et~al.}, ``{Damping signatures at
  JUNO, a medium-baseline reactor neutrino oscillation experiment},''
  \href{http://dx.doi.org/10.1007/JHEP06(2022)062}{{\em JHEP} {\bfseries 06}
  (2022) 062}, \href{http://arxiv.org/abs/2112.14450}{{\ttfamily
  arXiv:2112.14450 [hep-ex]}}.

\bibitem{Barenboim:2024wdn}
G.~Barenboim, A.~M. Calatayud-Cadenillas, A.~M. Gago, and C.~A. Ternes,
  ``{Quantum decoherence effects on precision measurements at DUNE and T2HK},''
  \href{http://dx.doi.org/10.1016/j.physletb.2024.138626}{{\em Phys. Lett. B}
  {\bfseries 852} (2024) 138626},
  \href{http://arxiv.org/abs/2402.16395}{{\ttfamily arXiv:2402.16395
  [hep-ph]}}.

\bibitem{KM3NeT:2024jji}
{\bfseries KM3NeT} Collaboration, S.~Aiello {\em et~al.}, ``{Search for quantum
  decoherence in neutrino oscillations with six detection units of
  KM3NeT/ORCA},'' \href{http://arxiv.org/abs/2410.01388}{{\ttfamily
  arXiv:2410.01388 [hep-ex]}}.

\bibitem{ICECUBE:2023gdv}
{\bfseries ICECUBE, IceCube} Collaboration, R.~Abbasi {\em et~al.}, ``{Search
  for decoherence from quantum gravity with atmospheric neutrinos},''
  \href{http://dx.doi.org/10.1038/s41567-024-02436-w}{{\em Nature Phys.}
  {\bfseries 20} no.~6, (2024) 913--920},
  \href{http://arxiv.org/abs/2308.00105}{{\ttfamily arXiv:2308.00105
  [hep-ex]}}.

\bibitem{DeRomeri:2023dht}
V.~De~Romeri, C.~Giunti, T.~Stuttard, and C.~A. Ternes, ``{Neutrino oscillation
  bounds on quantum decoherence},''
  \href{http://dx.doi.org/10.1007/JHEP09(2023)097}{{\em JHEP} {\bfseries 09}
  (2023) 097}, \href{http://arxiv.org/abs/2306.14699}{{\ttfamily
  arXiv:2306.14699 [hep-ph]}}.

\bibitem{Kamiokande-II:1987idp}
{\bfseries Kamiokande-II} Collaboration, K.~Hirata {\em et~al.}, ``{Observation
  of a Neutrino Burst from the Supernova SN 1987a},''
  \href{http://dx.doi.org/10.1103/PhysRevLett.58.1490}{{\em Phys. Rev. Lett.}
  {\bfseries 58} (1987) 1490--1493}.

\bibitem{Alekseev:1988gp}
E.~N. Alekseev, L.~N. Alekseeva, I.~V. Krivosheina, and V.~I. Volchenko,
  ``{Detection of the Neutrino Signal From {SN1987A} in the {LMC} Using the Inr
  Baksan Underground Scintillation Telescope},''
  \href{http://dx.doi.org/10.1016/0370-2693(88)91651-6}{{\em Phys. Lett. B}
  {\bfseries 205} (1988) 209--214}.

\bibitem{Bionta:1987qt}
R.~M. Bionta {\em et~al.}, ``{Observation of a Neutrino Burst in Coincidence
  with Supernova SN 1987a in the Large Magellanic Cloud},''
  \href{http://dx.doi.org/10.1103/PhysRevLett.58.1494}{{\em Phys. Rev. Lett.}
  {\bfseries 58} (1987) 1494}.

\bibitem{Lindblad:1975ef}
G.~Lindblad, ``{On the Generators of Quantum Dynamical Semigroups},''
  \href{http://dx.doi.org/10.1007/BF01608499}{{\em Commun. Math. Phys.}
  {\bfseries 48} (1976) 119}.

\bibitem{Gorini:1975nb}
V.~Gorini, A.~Kossakowski, and E.~C.~G. Sudarshan, ``{Completely Positive
  Dynamical Semigroups of N Level Systems},''
  \href{http://dx.doi.org/10.1063/1.522979}{{\em J. Math. Phys.} {\bfseries 17}
  (1976) 821}.

\bibitem{Gago:2002na}
A.~M. Gago, E.~M. Santos, W.~J.~C. Teves, and R.~Zukanovich~Funchal, ``{A Study
  on quantum decoherence phenomena with three generations of neutrinos},''
  \href{http://arxiv.org/abs/hep-ph/0208166}{{\ttfamily arXiv:hep-ph/0208166}}.

\bibitem{Domi:2024ypm}
A.~Domi, T.~Eberl, M.~J. Fahn, K.~Giesel, L.~Hennig, U.~Katz, R.~Kemper, and
  M.~Kobler, ``{Understanding gravitationally induced decoherence parameters in
  neutrino oscillations using a microscopic quantum mechanical model},''
  \href{http://dx.doi.org/10.1088/1475-7516/2024/11/006}{{\em JCAP} {\bfseries
  11} (2024) 006}, \href{http://arxiv.org/abs/2403.03106}{{\ttfamily
  arXiv:2403.03106 [gr-qc]}}.

\bibitem{DEsposito:2023psn}
V.~D'Esposito and G.~Gubitosi, ``{Constraints on quantum spacetime-induced
  decoherence from neutrino oscillations},''
  \href{http://dx.doi.org/10.1103/PhysRevD.110.026004}{{\em Phys. Rev. D}
  {\bfseries 110} no.~2, (2024) 026004},
  \href{http://arxiv.org/abs/2306.14778}{{\ttfamily arXiv:2306.14778
  [hep-ph]}}.

\bibitem{Nandi:2024wkm}
P.~Nandi and T.~Bhattacharyya, ``{Unearthing Neutrino Decoherence from Quantum
  Spacetime: An Open Quantum Systems Perspective},''
  \href{http://arxiv.org/abs/2410.03808}{{\ttfamily arXiv:2410.03808
  [hep-ph]}}.

\bibitem{Domi:2024muy}
A.~Domi, T.~Eberl, D.~Hellmann, S.~Krieg, and H.~P\"as, ``{Potential of
  neutrino telescopes to detect quantum gravity-induced decoherence in the
  presence of dark fermions},''
  \href{http://dx.doi.org/10.1088/1475-7516/2025/01/063}{{\em JCAP} {\bfseries
  01} (2025) 063}, \href{http://arxiv.org/abs/2409.12633}{{\ttfamily
  arXiv:2409.12633 [hep-ph]}}.

\bibitem{dosSantos:2023skk}
M.~V. dos Santos, P.~C. de~Holanda, P.~Dedin~Neto, and E.~Kemp, ``{Effects of
  quantum decoherence in a future supernova neutrino detection},''
  \href{http://dx.doi.org/10.1103/PhysRevD.108.103032}{{\em Phys. Rev. D}
  {\bfseries 108} no.~10, (2023) 103032},
  \href{http://arxiv.org/abs/2306.17591}{{\ttfamily arXiv:2306.17591
  [hep-ph]}}.

\bibitem{Martinez-Mirave:2024hfd}
P.~Mart\'\i{}nez-Mirav\'e, I.~Tamborra, and M.~T\'ortola, ``{The Sun and
  core-collapse supernovae are leading probes of the neutrino lifetime},''
  \href{http://dx.doi.org/10.1088/1475-7516/2024/05/002}{{\em JCAP} {\bfseries
  05} (2024) 002}, \href{http://arxiv.org/abs/2402.00116}{{\ttfamily
  arXiv:2402.00116 [astro-ph.HE]}}.

\bibitem{Ivanez-Ballesteros:2023lqa}
P.~Iv\'a\~nez Ballesteros and M.~C. Volpe, ``{SN1987A and neutrino
  non-radiative decay},''
  \href{http://dx.doi.org/10.1016/j.physletb.2023.138252}{{\em Phys. Lett. B}
  {\bfseries 847} (2023) 138252},
  \href{http://arxiv.org/abs/2307.03549}{{\ttfamily arXiv:2307.03549
  [hep-ph]}}.

\bibitem{Fiorillo:2023frv}
D.~F.~G. Fiorillo, M.~Heinlein, H.-T. Janka, G.~Raffelt, E.~Vitagliano, and
  R.~Bollig, ``{Supernova simulations confront SN 1987A neutrinos},''
  \href{http://dx.doi.org/10.1103/PhysRevD.108.083040}{{\em Phys. Rev. D}
  {\bfseries 108} no.~8, (2023) 083040},
  \href{http://arxiv.org/abs/2308.01403}{{\ttfamily arXiv:2308.01403
  [astro-ph.HE]}}.

\bibitem{Pagliaroli:2008ur}
G.~Pagliaroli, F.~Vissani, M.~L. Costantini, and A.~Ianni, ``{Improved analysis
  of SN1987A antineutrino events},''
  \href{http://dx.doi.org/10.1016/j.astropartphys.2008.12.010}{{\em Astropart.
  Phys.} {\bfseries 31} (2009) 163--176},
  \href{http://arxiv.org/abs/0810.0466}{{\ttfamily arXiv:0810.0466
  [astro-ph]}}.

\bibitem{Strumia:2003zx}
A.~Strumia and F.~Vissani, ``{Precise quasielastic neutrino/nucleon
  cross-section},'' \href{http://dx.doi.org/10.1016/S0370-2693(03)00616-6}{{\em
  Phys. Lett. B} {\bfseries 564} (2003) 42--54},
  \href{http://arxiv.org/abs/astro-ph/0302055}{{\ttfamily
  arXiv:astro-ph/0302055}}.

\bibitem{Loredo:2001rx}
T.~J. Loredo and D.~Q. Lamb, ``{Bayesian analysis of neutrinos observed from
  supernova SN-1987A},''
  \href{http://dx.doi.org/10.1103/PhysRevD.65.063002}{{\em Phys. Rev. D}
  {\bfseries 65} (2002) 063002},
  \href{http://arxiv.org/abs/astro-ph/0107260}{{\ttfamily
  arXiv:astro-ph/0107260}}.

\bibitem{Vissani:2014doa}
F.~Vissani, ``{Comparative analysis of SN1987A antineutrino fluence},''
  \href{http://dx.doi.org/10.1088/0954-3899/42/1/013001}{{\em J. Phys. G}
  {\bfseries 42} (2015) 013001},
  \href{http://arxiv.org/abs/1409.4710}{{\ttfamily arXiv:1409.4710
  [astro-ph.HE]}}.

\bibitem{deSalas:2020pgw}
P.~F. de~Salas, D.~V. Forero, S.~Gariazzo, P.~Mart\'\i{}nez-Mirav\'e, O.~Mena,
  C.~A. Ternes, M.~T\'ortola, and J.~W.~F. Valle, ``{2020 global reassessment
  of the neutrino oscillation picture},''
  \href{http://dx.doi.org/10.1007/JHEP02(2021)071}{{\em JHEP} {\bfseries 02}
  (2021) 071}, \href{http://arxiv.org/abs/2006.11237}{{\ttfamily
  arXiv:2006.11237 [hep-ph]}}.

\bibitem{Gariazzo:2022ahe}
S.~Gariazzo {\em et~al.}, ``{Neutrino mass and mass ordering: no conclusive
  evidence for normal ordering},''
  \href{http://dx.doi.org/10.1088/1475-7516/2022/10/010}{{\em JCAP} {\bfseries
  10} (2022) 010}, \href{http://arxiv.org/abs/2205.02195}{{\ttfamily
  arXiv:2205.02195 [hep-ph]}}.

\bibitem{Capozzi:2021fjo}
F.~Capozzi, E.~Di~Valentino, E.~Lisi, A.~Marrone, A.~Melchiorri, and
  A.~Palazzo, ``{Unfinished fabric of the three neutrino paradigm},''
  \href{http://dx.doi.org/10.1103/PhysRevD.104.083031}{{\em Phys. Rev. D}
  {\bfseries 104} no.~8, (2021) 083031},
  \href{http://arxiv.org/abs/2107.00532}{{\ttfamily arXiv:2107.00532
  [hep-ph]}}.

\bibitem{Esteban:2024eli}
I.~Esteban, M.~C. Gonzalez-Garcia, M.~Maltoni, I.~Martinez-Soler, J.~a.~P.
  Pinheiro, and T.~Schwetz, ``{NuFit-6.0: updated global analysis of
  three-flavor neutrino oscillations},''
  \href{http://dx.doi.org/10.1007/JHEP12(2024)216}{{\em JHEP} {\bfseries 12}
  (2024) 216}, \href{http://arxiv.org/abs/2410.05380}{{\ttfamily
  arXiv:2410.05380 [hep-ph]}}.

\bibitem{Janka:2025tvf}
H.~T. Janka, ``{Long-Term Multidimensional Models of Core-Collapse Supernovae:
  Progress and Challenges},'' \href{http://arxiv.org/abs/2502.14836}{{\ttfamily
  arXiv:2502.14836 [astro-ph.HE]}}.

\bibitem{Garching}
``Garching core-collapse supernova research archive,''
  \href{http://arxiv.org/abs/https://wwwmpa.mpa-garching.mpg.de/ccsnarchive/}{{\ttfamily
  https://wwwmpa.mpa-garching.mpg.de/ccsnarchive/}}.

\end{thebibliography}\endgroup

\end{document}